# Rejection of Smooth GPS Time Synchronization Attacks via Sparse Techniques

Erick Schmidt, *IEEE*, Junhwan Lee, Nikolaos Gatsis, *Member, IEEE,* and David Akopian, *Senior Member, IEEE*

*Abstract*—This paper presents a novel time synchronization attack (TSA) model for the Global Positioning System (GPS) based on clock data behavior changes in a higher-order derivative domain. Further, the time synchronization attack rejection and mitigation based on sparse domain (TSARM-S) is presented. TSAs affect stationary GPS receivers in applications where precise timing is required, such as cellular communications, financial transactions, and monitoring of the electric power grid. In the present work, the clock bias and clock drift are monitored at higher-order clock data derivatives where the TSA is seen as a sparse spike-like event. The smoothness of the attack relates to the derivative order where the sparsity is observed. The proposed method jointly estimates a dynamic solution for GPS timing and rejects behavior changes based on such sparse events. An evaluation procedure is presented for two testbeds, namely a commercial receiver and a software-defined radio. Further, the proposed method is evaluated against distinct real-dataset Texas Spoofing Test Battery (TEXBAT) scenarios. Combined synthetic and real-data results show an average RMS clock bias error of 12.08 m for the SDR platform, and 45.74 m for the commercial device. Further, the technique is evaluated against state-of-the-art mitigation techniques and in a spoofing-plus-multipath scenario for robustness. Finally, TSARM-S can be potentially optimized and implemented in commercial devices via a firmware upgrade.

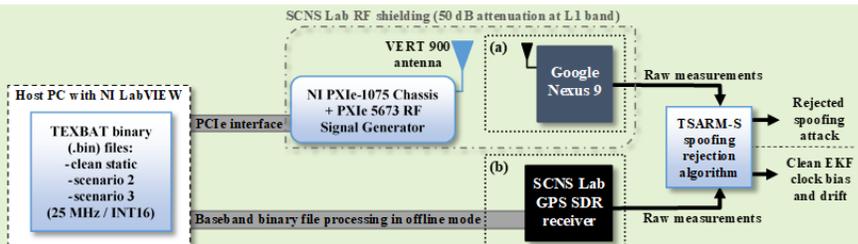

## I. Introduction

THE convergence of radionavigation systems, such as the Global Positioning System (GPS), with diverse domain applications, such as emergency response, air traffic control, financial transactions, and smart power grids, manifests their thriving popularity and availability [1]. However, due to its open accessibility for civilian use, GPS coarse acquisition (C/A) codes are subject to malicious cyber-physical attacks [2].

GPS cyber-physical attacks have been categorized into jamming and spoofing. While jamming completely blocks authentic signals via high power noise, spoofing exploits smart counterfeit signals to deceive and highjack a target receiver [3]. Once highjacked, the spoofer can inflict an erroneous position, velocity, and time (PVT) solution. Specifically, a Time Synchronization Attack (TSA) is a spoofing attack that alters the target receiver's clock offset. In particular, TSA spoofing can disrupt the operation of smart-grids which use Phasor Measurement Units (PMUs) that provide sensor readings precisely synchronized using GPS receivers [4]. An imprecise clock offset measurement due to a cyber-physical attack would cause stability control failures and power outages [5], [6]. Under a successful TSA, smart grids become exposed to transmission line faults, voltage instability, and missed event locations [7].

Authors in [2] categorize spoofing attacks based on their complexity, i.e., simplistic, intermediate, and advanced. The simplistic attack relies on retransmission of a delayed GPS signal with augmented power to inflict time delays. Intermediate attack uses a receiver-spoofer device that is placed near the target receiver, to retransmit smart GPS-like signals in a more covert fashion. Advanced spoofing uses several spoofer devices orchestrating a more elaborated synchronized attack. The intermediate attack is demonstrated as the most cost-effective in terms of implementation and can effortlessly synthesize a TSA. Additionally, cost-accessible software-defined radio (SDR) spoofers have successfully carried out attacks in such receivers as in [8] and [9]. This further rises awareness to harden receivers' abilities for spoofing attacks since conventional off-the-shelf receivers typically lack intermediate-to-advanced spoofing mitigation techniques [2].

Much has been discussed about monitoring and detection of spoofing attacks [10]. However, much work is still to be done on spoofing rejection and mitigation. Existing techniques often offer detection of spoofing or jamming, but lack an actual correction countermeasure. A countermeasure should be able to

This manuscript was first submitted to the IEEE Sensors Journal on April 3, 2020.
This work was supported by the National Science Foundation under Grant ECCS-1719043.

E. Schmidt, J. Lee, N. Gatsis and D. Akopian are with the Department of Electrical and Computer Engineering, The University of Texas at San Antonio, San Antonio, TX, 78249 USA (e-mail: erickschmidt@gmail.com; junhwan.lee@utsa.edu; nikolaos.gatsis@utsa.edu; david.akopian@utsa.edu).

detect and correct the attack while maintaining normal operation of the receiver with authentic PVT outputs. In particular, mitigation strategies can be organized in four categories [2]: (1) Advanced signal-processing-based techniques for standard single-antenna GNSS receiver that rely on power and automatic gain control (AGC) [11], complex correlation function outputs [12], conventional tracking loop [9], and vector-tracking loops (VTL) [13], [14]; (2) encryption-based defenses relying on encrypted GNSS signal keys that share a unique relationship between civilian open-access and military signals for spoofing detection [15]; (3) drift monitoring techniques that rely on unusual behavior changes in receiver position or clock via receiver observables [16], [17], positioning filter metrics [18], and sparse techniques [19], [20], [21]; and (4) signal-geometry-based defenses which rely on multi-antenna systems for angle-of-arrival spoofer detection and mitigation [22], [23]. For further categorization of spoofing countermeasures, the reader is directed to [2], [3], and [10].

This paper addresses the aforementioned lack of spoofing countermeasures and cost-effectiveness by introducing a technique that can be implemented by means of an inexpensive firmware upgrade to a GPS receiver. The proposed method relies on a dynamic model of the clock bias and clock drift for a stationary receiver. It specifically falls in the drift monitoring mitigation category as mentioned above [2]. Moreover, we focus on a single-antenna single-receiver architecture as opposed to complex multi-receiver architectures [13], [24], [25].

### A. Related Work

Different from previous works that detect and mitigate integrity anomalies in select parallel receiver channels [20], [21], [17], this work mitigates TSA interference even in the absence of integrity anomalies, when the smarter spoofer manipulates all the channels synchronously. The work in [17] relies on advanced receiver autonomous integrity monitoring (RAIM) techniques to detect anomalies per channel, and on every PVT computation. In fact, most RAIM-based techniques rely on anomaly checking per iteration, which can provide expensive computations.

In terms of dynamic modeling, the work in [18] monitor the Kalman Filter innovations for potential spoofer attacks, thus relying on simple metrics separately averaged from sequential data in a sliding window. Similarly, the work in [19] jointly computes a dynamic PVT solution and accumulates variation metrics, which are then accrued to gather the correction. They both rely on small spoofer alterations accumulated over time. On the other hand, the proposed method jointly computes the PVT solution while observing higher-order derivatives where the behavior change is detected in a sparse-domain, i.e., a small behavior change in the dynamic trend appears as a spike. The method automatically rejects behavior changes and outputs the estimated PVT solution without a need for correction.

Sparse techniques are applied on the pseudorange residuals for every PVT computation to detect anomalies in the measurements in [20]. Similarly, the work in [21] follows multi-frequency observables for enhanced outlier detection. An important assumption in such sparse estimation is that a small number of the visible satellites are corrupted, which entails sparsity in the measurement residual domain. While [20] uses sparse processing for outlier detection as an indication of integrity failure, the present work exploits sparsity for receiver behavior change detection and mitigation for all visible satellites. In other words, the proposed technique relies on sequential data observations for behavior change detection, as opposed to snapshot monitoring, and is applicable even when all visible satellites are simultaneously and consistently spoofed.

Finally, many state-of-the-art techniques based on sparse estimation focus on multipath (MP) detection and mitigation [20], [21]. While MP is similar to spoofing attacks, some differences are worth noting: (a) MP effects appear arbitrarily while spoofing is more consistent over time, (b) MP generally affects some satellite channels while spoofing attacks affect all channels concurrently, and (c) spoofing attacks are two orders of magnitude more hazardous in terms of PVT deviations, e.g., a meaningful TSA attack may inflict 8000 m, or equivalently, 26.67 µs bias error in the PMU clock [26].

### B. Contributions

Previous works have attempted to classify TSAs based on how smooth (or abrupt) the spoofer attack is induced onto the receiver clock offset [27]; however, such terms necessitate a refined categorization. In this work, we define the smoothness of the attack based on estimated clock data sequence change. The change is analyzed using clock bias derivative domains of various orders to detect the attack as a sparse spike-like event. Based on the sparse derivative domain, we categorize the attacks by their order, e.g., a *third order attack*.

The previously reported TSA attacks can be detected using the *third order attack* model of this paper. In particular, authors in [19] defined Type I and II attack models, which can be addressed using the proposed technique while achieving better mitigation results. This work develops a mitigation framework for up to *third order attacks*, but the proposed TSA framework extends to higher orders. Similarly, the proposed technique can detect other spoofing attacks as in [28], as will be seen in Section VI-B. In the following, the technique of this paper is referred to as time synchronization attack rejection and mitigation based on sparse-domain (TSARM-S).

The contributions of this paper are as follows:

1) Novel modeling of TSAs based on behavioral change analysis in clock data derivative domains of various orders is introduced by observing the derivative order in which sparsity shows up. The model allows for the attack to preserve its measurement integrity, which renders it undetected by traditional RAIM techniques.

2) Based on 1), we propose a dynamic model that jointly estimates the clock bias and drift, and rejects a potential TSA by transforming state variables into a higher-order sparse-domain where the TSA is detected and rejected.

3) The proposed model identifies and rejects the spoofing signatures in the clock bias and drift directly without the need to correct the authentic outputs, i.e., it splits the estimated clock outputs into authentic and spoofed.

4) The method is tested using real data corresponding to raw measurement outputs from two platforms: an in-house SDR from UTSA [29], and a Google Nexus 9

Tablet. Specifically, synthetic attacks are applied to real data recordings, and in addition, two real-data scenarios from The Texas Spoofing Test Battery (TEXBAT) are replayed over-the-air (OTA). Furthermore, a comparison with a spoofing plus MP scenario is evaluated.

The paper is organized as follows. Section II introduces a GPS dynamic model. Section III presents a novel TSA modeling. Section IV presents the proposed spoofing mitigation technique, TSARM-S. A testing methodology is described in Section V. Section VI presents simulation results and discussion. Finally, Section VII concludes and discusses future work.

## II. GPS PVT Dynamic Model

In this section, we briefly describe the radionavigation method used in GPS. To resolve the user's position, the GPS receiver uses satellite ranging signals which also contain satellite orbit parameters such as Ephemeris data to estimate the satellites' positions during location estimation [30]. The satellites serve as beacons for trilateration using satellite-to-user ranges measured by the receiver. Without the loss of generality, the user (GPS receiver) position can be represented in 3D Earth-centered Earth-fixed (ECEF) coordinates as $\mathbf{p}_u = \left[ x_u, y_u, z_u \right]^T$. Similarly, the position of the $n$-th satellite for $n = 1, 2, \ldots, N$ during each satellite transmit time $t_n$ is represented as $\mathbf{p}_n = \left[ x_n(t_n), y_n(t_n), z_n(t_n) \right]^T$. Further, we define the receive time at the receiver as $t_R$. The true range between user and the satellite can be defined as $d_n = \| \mathbf{p}_n - \mathbf{p}_u \|_2$, where $\| \|_2$ denotes the $\ell_2$-norm. However, the range is not known and can be expressed as the difference of the transmit and receive time as $d_n = c \left( t_R^{GPS} - t_n^{GPS} \right)$, where $t_n^{GPS}$ and $t_R^{GPS}$ are the accurate transmit and receive times, respectively. By introducing an offset in the measured user time of reception modeling the receiver clock inaccuracy as $t_R = t_R^{GPS} + b_u$, and likewise for the satellite transmit time as $t_n = t_n^{GPS} + b_n$, the receiver computes biased ranges called pseudoranges given by $\rho_n = c \left( t_R - t_n \right)$, where $c$ is the speed of light. One can rewrite the pseudorange equation by using the previous two definitions of $d_n$ as

$$\rho_n = \| \mathbf{p}_n - \mathbf{p}_u \|_2 + c \left( b_u - b_n \right) + \epsilon_{\rho_n} \quad (1)$$

where $\mathbf{p}_n$ is the satellite position at transmit time, $\mathbf{p}_u$ is the user position at receive time, $b_u$ and $b_n$ are the user and satellite clock offsets (in s), respectively, and $\epsilon_{\rho_n}$ models combined errors due to atmospheric delays, thermal noise, etc. (in m). The pseudoranges, satellite locations, and satellite clock offsets are known at and/or computed by the receiver, while $\left( \mathbf{p}_u, b_u \right)$ are estimated using (1).

Similarly, the receiver can measure the Doppler frequency shift (residual) that is formed on top of the L1 carrier frequency due to the relative difference between the satellite velocity, $\mathbf{v}_n$, and the user velocity, $\mathbf{v}_u$, also expressed in 3D ECEF coordinates. This estimated Doppler residual is related to the rate at which the pseudorange measurement varies over time, denoted as $\dot{\rho}_n$. The pseudorange rate is then represented as:

$$\dot{\rho}_n = \left( \mathbf{v}_n - \mathbf{v}_u \right)^T \frac{\mathbf{p}_n - \mathbf{p}_u}{\| \mathbf{p}_n - \mathbf{p}_u \|} + c \left( \dot{b}_u - \dot{b}_n \right) + \epsilon_{\dot{\rho}_n} \quad (2)$$

where $\mathbf{v}_n$ is the satellite velocity obtained from the navigation message, $\mathbf{v}_u$ is the user velocity, $\dot{b}_u$ and $\dot{b}_n$ are the user and satellite clock drifts (in s/s), and $\epsilon_{\dot{\rho}_n}$ is the modeled noise. Similarly to (1), the unknowns to be estimated from (2) are $\left( \mathbf{v}_u, \dot{b}_u \right)$.

For a conventional low-dynamics receiver, the PVT solution aims to solve for user position, velocity, and the receiver's clock bias and clock drift. This totals 8 unknown variables which are typically computed via Weighted Least Squares (WLS) in a snapshot manner [31], or dynamically by means of an Extended Kalman filter (EKF). The dynamic equation of an 8-state EKF amounts to a random walk model as follows [32]:

$$\mathbf{x}_k = \underbrace{\begin{pmatrix} \mathbf{\Phi} & \mathbf{0} & \mathbf{0} & \mathbf{0} \\ \mathbf{0} & \mathbf{\Phi} & \mathbf{0} & \mathbf{0} \\ \mathbf{0} & \mathbf{0} & \mathbf{\Phi} & \mathbf{0} \\ \mathbf{0} & \mathbf{0} & \mathbf{0} & \mathbf{\Phi} \end{pmatrix}}_{\mathbf{F}_k} \mathbf{x}_{k-1} + \mathbf{w}_k \quad (3)$$

where $\mathbf{x} \equiv \left[ x_u \ \dot{x}_u \ y_u \ \dot{y}_u \ z_u \ \dot{z}_u \ cb_u \ c\dot{b}_u \right]^T$ is the state vector, $cb_u$ and $c\dot{b}_u$ are the user clock bias (in m) and clock drift (in m/s), $\mathbf{p}_u = \left[ x_u, y_u, z_u \right]^T$ is the user location where the components are in meters (m), $\mathbf{v}_u = \left[ \dot{x}_u, \dot{y}_u, \dot{z}_u \right]^T$ denotes user velocity in m/s, $\mathbf{w}_k$ is the process noise, and $\mathbf{\Phi}$ is a state-transition matrix for the discrete time instance $k$ corresponding to each position-velocity pair as follows:

$$\mathbf{\Phi} = \begin{bmatrix} 1 & \Delta t \\ 0 & 1 \end{bmatrix} \quad (4)$$

where $\Delta t$ is the discretization time interval for each measurement. Lastly, the measurements given by (1) and (2) for pseudoranges and pseudorange rates are used as inputs to the 8-state EKF based on (3) for the dynamic PVT solution. Note that (1) and (2) model different observables from different circuitry sources of the receiver, which are respectively extracted from code-phases and Doppler residuals measured by the receiver [30]; however, they are used jointly for the navigation computation.

## III. Novel TSA Modeling

In this section, we present a novel modeling for TSAs that covers a wide range of attacks. The proposed concept interprets smooth attacks as the receiver's clock dynamic behavior change. The change is detected by inspecting higher order derivatives of the sequence of estimated clock data. It is demonstrated that the TSA manifests itself as a sparse event such as a combination of few spikes at one of the derivative clock signals. The smoother the attack, the higher-order derivative is required to detect the sparse indication of behavior change. Thus, the TSAs are systematized based on such higher-order clock signal derivative domains. We begin by listing some self-consistent spoofer requirements [2].

### A. Measurement integrity checks

In this subsection, we define two measurement integrity checks associated with the previously defined dynamic model. We assume these integrity checks are incorporated by the spoofer attacks to avoid detection using straightforward techniques.

The attack on pseudoranges and pseudorange rates is modeled as follows:

$$\begin{aligned}\rho_{n,s}[k] &= \rho_n[k] + s_\rho[k] \\ \dot{\rho}_{n,s}[k] &= \dot{\rho}_n[k] + s_{\dot{\rho}}[k]\end{aligned} \quad (5)$$

where $\rho_{n,s}$ and $\dot{\rho}_{n,s}$ are the spoofed pseudorange (in m) and pseudorange rate (in m/s) measurements for the $n$-th satellite, and $s_\rho$ and $s_{\dot{\rho}}$ are the spoofing alterations on pseudoranges and pseudorange rates, respectively. TSAs attempt to steer the user clock bias and clock drift without altering the user position and velocity. To achieve this, the spoofer alterations $s_\rho$ and $s_{\dot{\rho}}$ must be the same in magnitude for all visible satellites. In this case, although the spoofer alterations are the attacks on pseudoranges and pseudorange rates, these attacks will be reflected on the clock bias and drift of the target receiver, respectively [19], [28]. This type of spoofer is categorized as an intermediate attack following [2], and will be an assumption throughout this work. These attacks also avoid rudimentary schemes that check measurement integrity, such as RAIM [33].

As stated in Section II, the measurement observables, $\rho_n$ and $\dot{\rho}_n$, come from different circuitry parts of the receiver; nonetheless, such measurements should have an integrity due to their physical interpretation. Thus, the first test to determine if measurement integrity is maintained between the pseudoranges and pseudorange rates is defined as follows:

$$\dot{\rho}_n[k] \approx \frac{\rho_n[k] - \rho_n[k-1]}{\Delta t} \quad (6)$$

Note that the derivative relationship in (6) also holds for the spoofed measurements in (5), namely $\rho_{n,s}$, and $\dot{\rho}_{n,s}$, as well as the spoofer alterations, $s_\rho$ and $s_{\dot{\rho}}$. In fact, this relationship of the spoofer alterations is assumed as part of the smartly devised attack.

Similarly, because TSAs reflect on the clock bias and clock drift after the PVT computation, the second integrity check is defined as follows:

$$\hat{c b_u}[k] \approx \frac{c\hat{b}_u[k] - c\hat{b}_u[k-1]}{\Delta t} \quad (7)$$

where $\hat{b}_u$ and $\hat{\dot{b}}_u$ are the estimated clock bias and drift produced by WLS.

By considering both (6) and (7) as part of a smart self-consistent TSA, the alterations $s_\rho$ and $s_{\dot{\rho}}$ are directly reflected in the clock bias and clock drift WLS outputs. Thus, without the loss of generality, the spoofer is assumed to perpetrate a TSA with two integrity considerations:

1) The alterations on pseudoranges and pseudorange rates are done on all visible channels, simultaneously, and with the same magnitude. This inflicts a deviation on the clock bias and clock drift while avoiding RAIM detection schemes.
2) The spoofing attack is performed while maintaining measurement integrity checks on (6) and (7).

To summarize, failure of the attack to satisfy (6) or (7) would be the basis for quick and straightforward detection. The next section focuses on the characteristics of smart attacks that will enable their rejection and mitigation, even when (6) and (7) are satisfied.

### B. Higher-order attacks

We define higher-order *derivatives* of the attack on the pseudorange, $s_\rho[k]$, to categorize the attacks according to the order where the attack appears as sparse. Table 1 lists higher order user clock data modeling for TSAs. Such categories are derived from a classical physical interpretation of an object displacement over time, where the clock bias corresponds to the position (in m), the clock drift is velocity (in m/s), etc. This should not be confused with GPS dynamic clock modeling such as in [34], and [35]; rather, we use such definitions to facilitate attack detection.

We define the following equations related to $s_\rho[k]$ for velocity, acceleration, and jerk attacks, respectively:

$$\begin{aligned}s_{\dot{\rho}}[k] &= \frac{s_\rho[k] - s_\rho[k-1]}{\Delta t} \\ s_{\ddot{\rho}}[k] &= \frac{s_{\dot{\rho}}[k] - s_{\dot{\rho}}[k-1]}{\Delta t} \\ s_{\dddot{\rho}}[k] &= \frac{s_{\ddot{\rho}}[k] - s_{\ddot{\rho}}[k-1]}{\Delta t}\end{aligned} \quad (8)$$

Further, the following categories are defined: a) *first order attack*, b) *second order attack*, and c) *third order attack*. A *first order attack* occurs when the sequence $s_{\dot{\rho}}[k]$ is sparse; a *second order attack* occurs when $s_{\ddot{\rho}}[k]$ is sparse, but not $s_{\dot{\rho}}[k]$; a third order attack occurs when $s_{\dddot{\rho}}[k]$ is sparse, but not $s_{\ddot{\rho}}[k]$. The attack appears increasingly smooth as the order of the attack is higher, e.g., a *third order attack* is smoother than a *second order attack*. These categories do not define how the spoofer attack is devised, rather they define the order where the sparse event occurs and enable detection and rejection of the attack, as will be developed in Section IV.

TABLE I
HIGHER ORDER USER CLOCK MODELING FOR TSAS

| TSA attack exhibits itself as a sparse event on | Attack (derivative) order | Units |
|---|---|---|
| Clock bias | 0th order | m |
| Clock drift (velocity) | 1st order | m/s |
| Clock acceleration | 2nd order | m/s² |
| Clock jerk | 3rd order | m/s³ |

Fig. 1 shows examples of *first, second,* and *third order attacks* from top to bottom. The *first order attack* (top) appears as a step function attack on the clock bias with its derivative being a sparse-peak at the clock drift. The *second order attack* appears as a peak on the clock acceleration, a step on the clock drift, and a (smooth) ramp on the clock bias. An even smoother clock bias is seen on the *third order attack,* where the sparse peak appears on the clock jerk. Finally, it can be seen that the highest order attack encompasses its lower-order attacks, e.g., the *first order attack* is also sparse on the third order. This can be useful in terms of detection.

Several attacks reported in literature are defined as Type I and II attacks [16], [19]. However, the proposed TSA modeling amounts to a broader framework that incorporates previous definitions. It can be seen in Fig. 1 that Type I attack is in fact a *first order attack*, and Type II attack is a *third order attack*. However, the *second order attack* is not previously defined. Table 2 shows the relationship of these previous definitions with the new TSA models.

In this analysis, the higher the order where the sparse spike-like events appear, the subtler the attack on the clock bias becomes. Hence, while the significance of the attack smoothness has been noted in the literature [27], a systematic definition of *smoothness* is formulated in the present work. It is also mentioned in the literature that the smoothness of the attack on the clock bias is relevant for phase measurement units (PMUs) in smart-grids [5], [6]. Finally, this TSA analysis can be effortlessly extended to higher orders, e.g., clock snap (fourth order), and crackle (fifth order). Based on the existing literature, the third order analysis (clock jerk) covers enough intermediate spoofing attacks for practical purposes.

## IV. TSA REJECTION VIA A SPARSE TECHNIQUE

This section introduces a joint dynamic model and $\ell_1$-minimization problem which incorporates TSA models up to a third-order derivative (clock jerk). The dynamic model introduces the spoofing attack in the measurement model. Additionally, it penalizes the outlier based on sparse-domain TSA models previously discussed in Section III-B.

### A. Dynamic model on User Clock

The dynamic model presented here pertains to a stationary receiver and assumes that the user position $\mathbf{p}_u$ is known and the user velocity $\mathbf{v}_u$ is zero [16], [19]. Thus, the model is simplified to estimate only the user clock bias and clock drift as follows:

$$\underbrace{\begin{pmatrix} cb_u[k] \\ \dot{cb}_u[k] \end{pmatrix}}_{\mathbf{x}_k} = \underbrace{\begin{pmatrix} 1 & \Delta t \\ 0 & 1 \end{pmatrix}}_{\mathbf{F}_k} \underbrace{\begin{pmatrix} cb_u[k-1] \\ \dot{cb}_u[k-1] \end{pmatrix}}_{\mathbf{x}_{k-1}} + \underbrace{\begin{pmatrix} cw_b[k] \\ cw_{\dot{b}}[k] \end{pmatrix}}_{\mathbf{w}_k} \quad (9)$$

where $\mathbf{x} \equiv \begin{bmatrix} cb_u & \dot{cb}_u \end{bmatrix}^T$ is the 2-state vector, $\mathbf{F}_k$ is the state transition matrix, and $\mathbf{w}_k$ is the process noise vector that follows a white Gaussian covariance matrix $\mathbf{Q}_k$ related to the crystal oscillator of the user receiver [31].

The spoofer alterations are introduced in the state vector to capture the state estimate along with a potential attack.

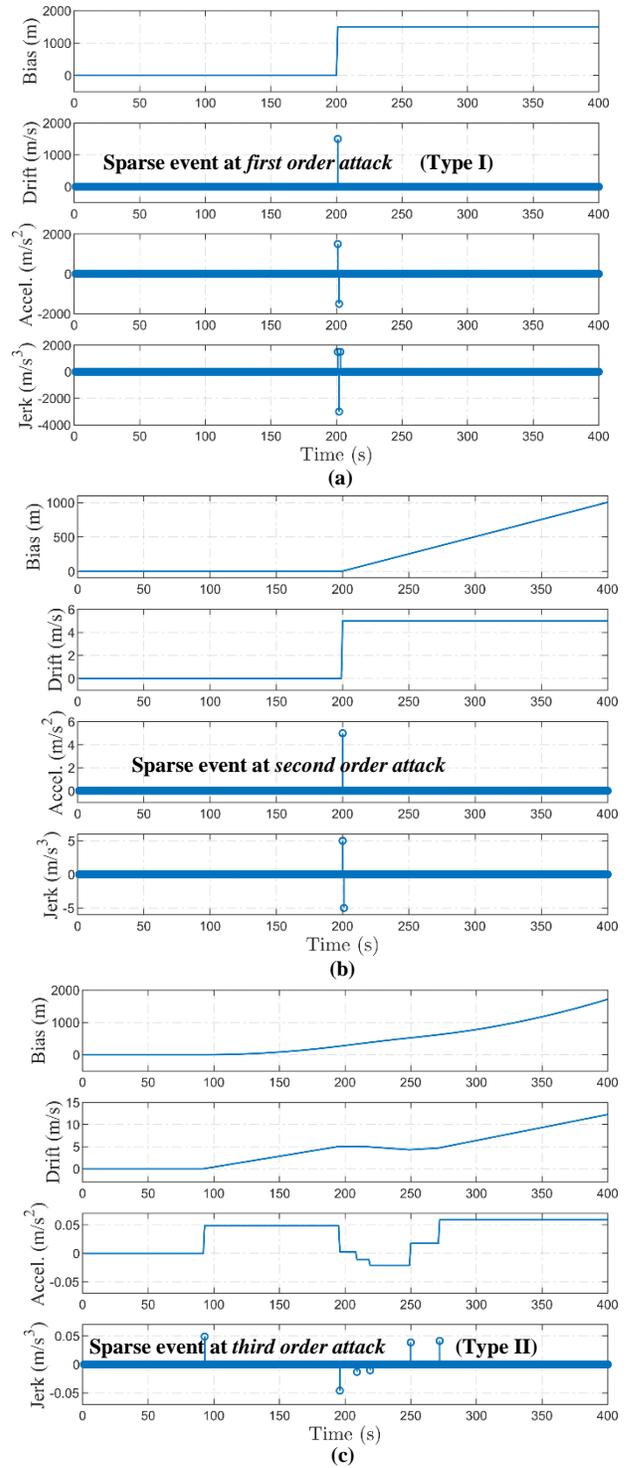

Fig. 1. TSA modeling for higher order derivative for (a) a *first order attack*, (b) a *second order attack*, and (c) a *third order attack.*.

Specifically, we define $\mathbf{s}_k = \begin{bmatrix} cs_b[k], cs_{\dot{b}}[k] \end{bmatrix}^T$ as the spoofer alteration vector, where $s_b$ and $s_{\dot{b}}$ are the attacks on the clock bias and clock drift, respectively. Based on assumptions from Section III.B, it holds that $s_\rho \equiv cs_b$ and $s_{\dot{\rho}} \equiv cs_{\dot{b}}$. Additionally, we define $\boldsymbol{\rho}[k] = \begin{bmatrix} \rho_1[k], ..., \rho_N[k] \end{bmatrix}^T$, and $\dot{\boldsymbol{\rho}}[k] = \begin{bmatrix} \dot{\rho}_1[k], ..., \dot{\rho}_N[k] \end{bmatrix}^T$, as the pseudorange and pseudorange rate measurement vectors, respectively. We then write the measurement equation by combining (1), (2), and (5) as follows:

$$
\underbrace{\begin{pmatrix} \boldsymbol{\rho}[k] \\ \dot{\boldsymbol{\rho}}[k] \end{pmatrix}}_{\mathbf{y}_k} = \underbrace{\begin{pmatrix} \mathbf{1}_{N\times 1} & \mathbf{0}_{N\times 1} \\ \mathbf{0}_{N\times 1} & \mathbf{1}_{N\times 1} \end{pmatrix}}_{\bar{\mathbf{H}}_k} \underbrace{\begin{pmatrix} cb_u[k] \\ c\dot{b}_u[k] \end{pmatrix}}_{\mathbf{x}_k} + \underbrace{\begin{pmatrix} \mathbf{1}_{N\times 1} & \mathbf{0}_{N\times 1} \\ \mathbf{0}_{N\times 1} & \mathbf{1}_{N\times 1} \end{pmatrix}}_{\bar{\mathbf{H}}_k} \underbrace{\begin{pmatrix} cs_b[k] \\ c\dot{s}_b[k] \end{pmatrix}}_{\mathbf{s}_k}
$$

$$
+ \underbrace{\begin{pmatrix} \|\mathbf{p}_1[k] - \mathbf{p}_u[k]\|_2 - cb_1[k] \\ \vdots \\ \|\mathbf{p}_N[k] - \mathbf{p}_u[k]\|_2 - cb_N[k] \\ (\mathbf{v}_1[k] - \mathbf{v}_u[k])^T \cdot \dfrac{\mathbf{p}_1[k] - \mathbf{p}_u[k]}{\|\mathbf{p}_1[k] - \mathbf{p}_u[k]\|_2} - c\dot{b}_1[k] \\ \vdots \\ (\mathbf{v}_N[k] - \mathbf{v}_u[k])^T \cdot \dfrac{\mathbf{p}_N[k] - \mathbf{p}_u[k]}{\|\mathbf{p}_N[k] - \mathbf{p}_u[k]\|_2} - c\dot{b}_N[k] \end{pmatrix}}_{\mathbf{c}_k} + \underbrace{\begin{pmatrix} \boldsymbol{\epsilon}_{\rho_1} \\ \vdots \\ \boldsymbol{\epsilon}_{\rho_N} \\ \boldsymbol{\epsilon}_{\dot{\rho}_1} \\ \vdots \\ \boldsymbol{\epsilon}_{\dot{\rho}_N} \end{pmatrix}}_{\boldsymbol{\epsilon}_k}
$$

$$(10)$$

where $\mathbf{c}_k$ is a known sequence that relates to the known user position, velocity (which is zero), satellite position, velocity and clock corrections, $\boldsymbol{\epsilon}_k$ is the zero mean Gaussian measurement noise with covariance matrix $\mathbf{R}_k = diag\left(\sigma_{\rho_1}^2[k], \dots, \sigma_{\rho_N}^2[k], \sigma_{\dot{\rho}_1}^2[k], \dots, \sigma_{\dot{\rho}_N}^2[k]\right)$, and $\sigma_{\rho_n}^2$ and $\sigma_{\dot{\rho}_n}^2$ are the uncorrelated measurement errors per satellite [31]. Equations (9) and (10) can be written as:

$$
\begin{aligned}
\mathbf{x}_k &= \mathbf{F}_k \mathbf{x}_{k-1} + \mathbf{w}_k \\
\mathbf{z}_k &= \mathbf{H}_k \mathbf{x}_k + \mathbf{H}_k \mathbf{s}_k + \boldsymbol{\epsilon}_k
\end{aligned}
\tag{11}
$$

where $\mathbf{z}_k = \mathbf{y}_k - \mathbf{c}_k$ is called the measurement residual.

TABLE II
TSA MODELING COMPARISON WITH PREVIOUS ATTACK TYPES [19]

| New attack category | Previously defined attack category | Shape |
|---|---|---|
| First order attack | Type I attack | Step |
| Second order attack | N/A | Ramp |
| Third order attack | Type II attack | Gradual |

### B. TSARM-S problem formulation

TSARM-S focuses on sparse-like behavior changes occurring on a higher-order derivative. To achieve this, we introduce an outlier detection scheme in the measurement model and define a $\ell_1$-minimization problem from the dynamic model in (11) for $k = 1, \dots, K$ measurements as follows:

$$
(\hat{\mathbf{x}}, \hat{\mathbf{s}}) = \underset{\mathbf{x}, \mathbf{s}}{\operatorname{argmin}} \left\{ \frac{1}{2} \sum_{k=1}^{K} \|\mathbf{z}_k - \mathbf{H}_k \mathbf{x}_k - \mathbf{H}_k \mathbf{s}_k\|_{\mathbf{R}_k^{-1}}^2 \right.
$$
$$
\left. + \frac{1}{2} \sum_{k=1}^{K} \|\mathbf{x}_k - \mathbf{F}_k \mathbf{x}_{k-1}\|_{\mathbf{Q}_k^{-1}}^2 + \lambda \|\mathbf{D}_2 \mathbf{s}'\|_{\ell_1} \right\}
\tag{12}
$$

where $\|\mathbf{x}\|_{\mathbf{M}}^2 = \mathbf{x}^T \mathbf{M} \mathbf{x}$, $\hat{\mathbf{x}} = [\hat{\mathbf{x}}_0, \dots, \hat{\mathbf{x}}_K]^T$ are the estimated states, $\hat{\mathbf{s}} = [\hat{\mathbf{s}}_1, \dots, \hat{\mathbf{s}}_K]^T$ are the estimated spoofer alterations,

$\mathbf{s}' = \left[cs_{\dot{b}}[1], \dots, cs_{\dot{b}}[k]\right]^T$ is a sub-vector of $\mathbf{s}$ which only contains the alterations on the clock drift, $\lambda$ is a tuning parameter, and $\mathbf{D}_2$ is a $K \times K$ second order total variation (TV) matrix applied to $K$ spoofer alterations of the clock drift $cs_{\dot{b}}$ in $\mathbf{s}'$ and is defined as follows [36]:

$$
\mathbf{D}_2 = \begin{pmatrix} -2 & 1 & 0 & \dots & 0 \\ 1 & -2 & 1 & \dots & 0 \\ \vdots & \ddots & \ddots & \ddots & \vdots \\ 0 & \dots & 1 & -2 & 1 \end{pmatrix}
\tag{13}
$$

The first term of (12) comes from the measurement equation which contains the state and attack estimates. At the same time, the second term defines the EKF random walk model, and finally the third term promotes sparsity on the jerk of the attack by applying a second order TV matrix on the attack velocity. If an attack is present, that is, if sparse peaks are found on the third-order derivative, the model rejects the clock alterations from the state vector $\hat{\mathbf{x}}$, and places them in vector $\hat{\mathbf{s}}$ instead. In other words, there is no need for a correction stage, as rejection occurs automatically, given the proper tuning of $\lambda$.

Additionally, the first and second term jointly constrain the measurement integrity in the user clock bias and clock drift, corresponding to equations (6) and (7). Also, if the measurement integrity does not hold for the dynamical model in (11), the alteration variables $s_b$ and $s_{\dot{b}}$ absorb such erratic behaviors. This means that the proposed model is able to capture the attack as either a measurement integrity failure, or as a sparse event regardless of the TSA. Therefore, the functionality of the proposed method is three-fold:

1) A dynamic model of the user clock data based on the first and second term of (12);
2) An outlier detection and automatic capturing based on the first and third term of (12), and based on the third degree clock jerk, TSA modeling from Section III.B, and the promoted sparsity; and
3) A measurement integrity detection that is absorbed by the alteration variables $s_b$ and $s_{\dot{b}}$.

The next section addresses the testing methodology and presents numerical results achieved from (12).

## V. TESTING METHODOLOGY ON TSAs

As an effort to study TSA procedures on real-time receivers, this section discusses a testing methodology and testbed used for the proposed TSARM-S technique. Recently, the Android Location Team from Google have made available GNSS raw measurements to study high accuracy positioning techniques relevant to mass market applications [37]. They provide an Android application *GNSS Logger* along with MATLAB post-processing scripts to obtain pseudoranges and pseudorange rates from select Android devices. This provides an opportunity for commercial device testing on well-known spoofing testbeds such as TEXBAT [28]. Also, increasing in popularity are real-time SDR solutions [29], which provide access to the receiver chain from baseband to the navigation domain.

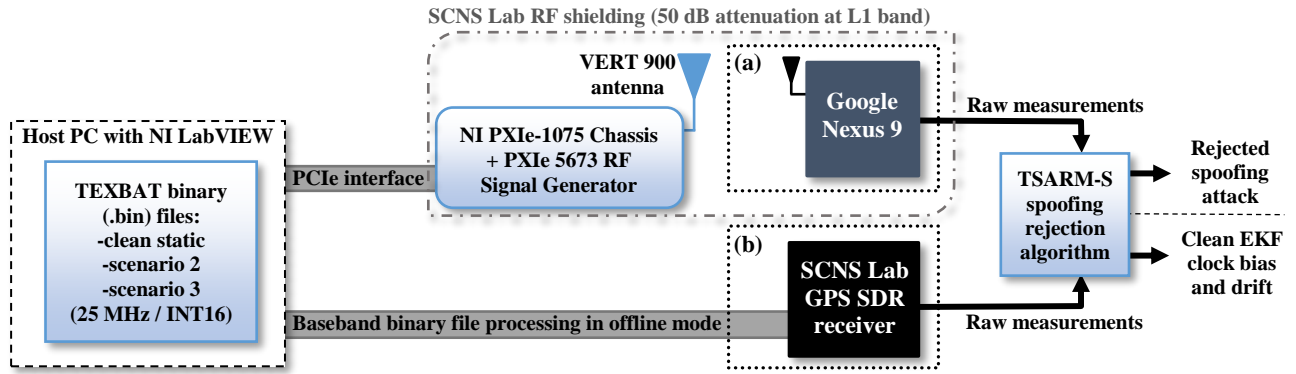

Fig. 2. Testbed for TEXBAT testing on TSARM-S with (a) record-and-replay on Google Nexus 9, and (b) baseband offline processing on a GPS SDR receiver.

### A. Testbed setup

In this work, we present a study on the effects of previously discussed TSAs (see Section III-B) on an integral testbed at the University of Texas at San Antonio's (UTSA) Software Communication & Navigation Systems (SCNS) Lab.

This study includes two main platforms for spoofing research along with real-data TEXBAT scenarios: (a) a commercial-grade Android-based Google Nexus 9 tablet with an embedded GPS chipset, providing raw measurements and MATLAB post-processing scripts from [37]; and (b) an in-house real-time LabVIEW-based single-frequency GPS L1 SDR receiver previously reported in [29], which provides raw measurements in a similar fashion. Three TEXBAT scenarios are explored, namely spoofing scenarios 2 and 3, and a clean static scenario for comparison purposes. The integral testbed for both testbed receivers (a) and (b) can be seen in Fig. 2.

We assess the testbed in two steps: first, we inject synthetic *first, second* and *third order TSA attacks* on pseudoranges and pseudorange rates for the TEXBAT clean static scenario; and second, we process TEXBAT scenarios 2 and 3 as real spoofing attacks. The synthetic simulations step provides a worst-case scenario where the attacks occur with negligible losses in carrier and code alignment, i.e., a perfect spoofing attack. The real spoofing attacks from TEXBAT scenarios 2 and 3 provide a more realistic setup and less far from a worst-case attack.

TEXBAT binary files are available at UT Austin Radionavigation Laboratory website [38]. For testbed (a), since the Nexus is a commercial receiver, we replay the TEXBAT recordings over-the-air (OTA) by using the following NI equipment: A host PC with a LabVIEW-based record-and-replay software, an NI PXIe-1075 Chassis with a PXIe 5673 RF Signal Generator via PCIe interface, and a VERT 900 antenna. For OTA transmissions, the SCNS Lab is equipped with a custom-made RF shielding area explicitly designed for GPS research. It uses 50 dB attenuation curtains at the L1 band to follow FCC regulations. The specifications of the TEXBAT recordings are at a 25 MHz sampling rate with INT16 in-phase and quadrature interleaved baseband samples, adequate for the NI equipment. The Nexus is properly shielded as to receive OTA replayed recordings. For (b), the binary files from TEXBAT are replayed in offline mode directly into the SDR receiver, thus avoiding OTA transmission effects.

### B. Synthetic simulations

The synthetic attack simulations are implemented on the clean static scenario.

| | Parameter | Value |
|---|---|---|
| TEXBAT scenarios 2 and 3 | Recording length (s) | 241 |
| | GPS week[a] | 1705 |
| | GPS sec[a] | 477986 to 478226 |
| | SVs for baseband processing | 3, 6, 10, 13, 16, 19, 23, 30 |
| First order attack | Attack start time (s) | 100 |
| | Bias max magnitude (m) | 1500 |
| | Drift max magnitude (m/s) | 1500 |
| Second order attack | Attack start time (s) | 50 |
| | Bias max magnitude (m) | 955 |
| | Drift max magnitude (m/s) | 5 |
| Third order attack | Attack start time (s) | 10 |
| | Bias max magnitude (m) | 750 |
| | Drift max magnitude (m/s) | 5 |

[a]The TEXBAT recording UTC Time is September 14, 2012 at 12:50:10.41 PM

Table 3 shows the recording length and specific GPS time used, along with synthetic attack parameters. The GPS time is used to synchronize between different TEXBAT scenarios. Also, all final attack bias magnitudes sufficiently surpass the distance of 600 m or 2 µs in time, for a complete channel capture [27]. Fig. 3 shows the synthetic *third order attack* injected to the pseudoranges and pseudorange rates to the clean static scenario as per the measurement integrity checks described in Section III-A. This shape is chosen to assimilate TEXBAT scenario 2, as will be seen in Section VI-B [28]. As for the synthetic *first* and *second order attacks*, see Fig. 1 shapes along with start and stop times seen in Table 3.

### C. Offline MATLAB evaluations

For mitigation, we implement TSARM-S in MATLAB environment in offline mode after obtaining the raw measurements from testbeds (a) and (b) (see Fig. 2). To calculate the measurement errors in the matrix $\mathbf{R}_k$, the Nexus post-processing scripts provide such values [37], and the SDR has its own implementation based on receiver characteristics [31], [32]. To model the clock error covariance matrix $\mathbf{Q}_k$, the Allan variance coefficients for a temperature-controlled crystal oscillator (TCXO) are used [32]. For the problem formulation, the MATLAB-based convex optimization solver CVX [39] is used. The output of a conventional EKF is used to obtain the *ground truth* from the TEXBAT clean static scenario for comparison against synthetic attacks, and TEXBAT scenarios 2 and 3. Regarding the $\lambda$ lambda parameter, Nexus simulations used values between 0.05 and 0.2, and the SDR utilized values

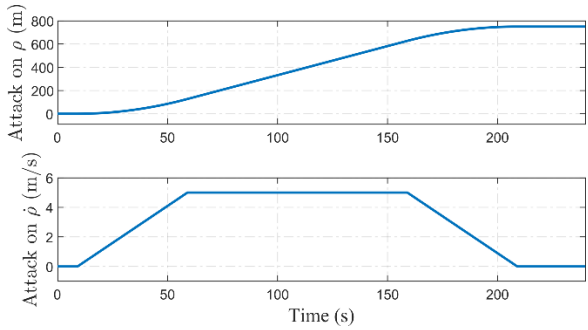

Fig. 3. Synthetic *third order attack* on pseudoranges and pseudorange rates.

between 250 and 1000. The tuning of this parameter can be achieved via cross-validation [39].

### D. Performance metric

As for performance metric, a root mean square error (RMSE) on the estimated clock bias is used for all scenarios. Let $K$ denote the total length of observation time. The RMSE is defined as:

$$RMSE = \sqrt{\frac{1}{K}\sum_{k=0}^{K-1}\left(c\hat{b}_u[k] - cb_{u,GT}[k]\right)^2} \qquad (14)$$

where $cb_{u,GT}$ is the *ground truth* clock bias, and $c\hat{b}_u$ is the estimated clock bias for each method.

### E. Scenarios

A total of nine scenarios are evaluated, as seen in Table 4. In the ensuing Section VI, only illustrative scenarios are shown to demonstrate the TSARM-S method on a commercial receiver, while the SDR testbed is used to further validate the results. The Nexus did not post-process scenario 3 properly because this attack required additional tuning and RAIM bypass on the SDR as will be discussed in Section VI-C. Additionally, a comparison of the clean static with a first order attack and with MP against state-of-the-art techniques is discussed in Sections VI-D and VI-E.

## VI. SIMULATION RESULTS

### A. Synthetic simulation results

This subsection presents results for synthetic attacks on the Nexus testbed. The Nexus results are shown initially to demonstrate TSARM-S capabilities on commercial receivers, and both Nexus and SDR results are shown in the next subsection for validation.

#### 1) Synthetic first order attack on Nexus 9

Fig. 4 shows the results for the TEXBAT clean static scenario with the *first order synthetic attack*. Fig. 4(a) shows the clock bias and clock drift for the clean, attacked, and corrected outputs of the EKF and TSARM-S, respectively. The clock bias is corrected to a 20.17 m RMSE while the drift is maintained at less than 1 m/s. Fig. 4(b) shows the estimated spoofer alteration values, $s_\rho$ and $s_{\dot\rho}$, obtained directly from the simulations. It can be seen that the attack is properly captured in the outlier vector and the sparse peaks are seen in the clock jerk.

TABLE IV
SUMMARY OF EVALUATED SCENARIOS

|  | TEXBAT clean static synthetic attacks | | | TEXBAT real scenarios | |
|  | 1st order | 2nd order | 3rd order | Scenario 2 | Scenario 3 |
|---|---|---|---|---|---|
| Nexus 9 | ✓ | ✓ | ✓ | ✓ | ✗ |
| SDR | ✓ | ✓ | ✓ | ✓ | ✓ |

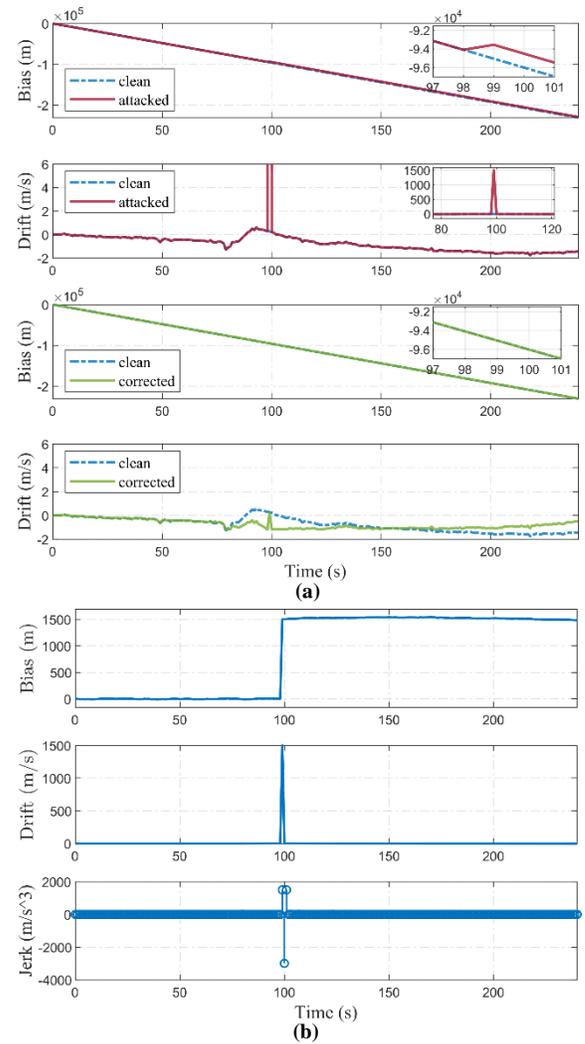

Fig. 4. Clean, attacked, and corrected clock bias and clock drift plots on (a), and estimated spoofer alterations on (b) for *first order attack* on Nexus 9.

#### 2) Synthetic second order attack on Nexus 9

Fig. 5 shows the results for the *second order attack* on the Nexus device. The characteristic ramp attack on the clean vs. attacked clock bias is smoother than the step attack, nonetheless, a total bias attack of almost 1000 m is seen in second 229 (see Fig. 5(a) zoom-in plot). The corrected bias shows an offset of 99 m bias at the same second. And the corrected clock drift is within 1 m/s. Further $\lambda$ tuning and proper clock modeling could improve this output, however the RMSE is quite an acceptable 42.47 m bias error from ground truth. Also, estimated spoofer alterations in Fig. 5(b) show clear ramp and step shapes detected for the clock bias and drift, respectively, and the clock jerk shows evident detection spikes.

#### 3) Synthetic third order attack on Nexus 9

The synthetic *third order attack* results on Nexus is seen in Fig 6. Out of all three, this is the subtlest and hardest to detect. In fact, it assimilates TEXBAT scenario 2 [28]. Nonetheless, Fig. 6(b) displays the clock jerk spikes that are reflected on the estimated spoofer bias and drift alterations. The RMSE of the corrected bias is 93.38 m, and the corrected clock drift is within 1 m/s as opposed to the injected 5 m/s attack.

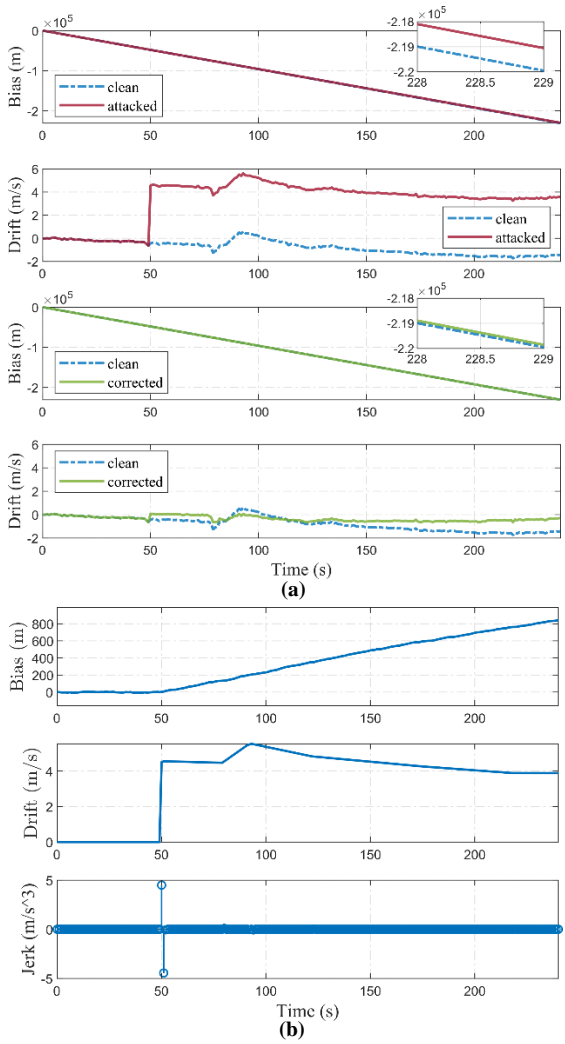

Fig. 5. Clean, attacked, and corrected clock bias and clock drift plots on (a), and estimated spoofer alterations on (b) for *second order attack* on Nexus 9.

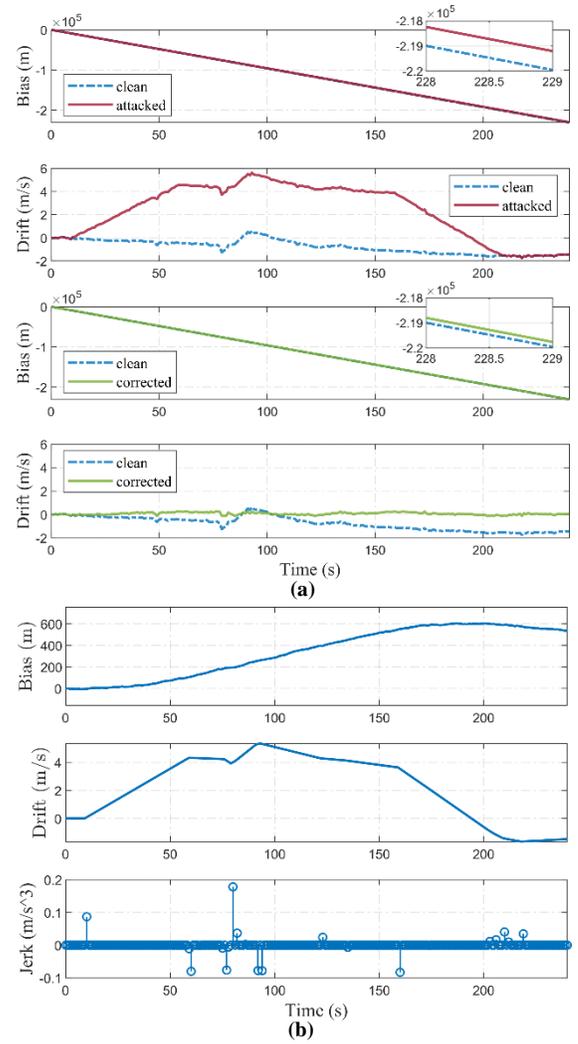

Fig. 6. Clean, attacked, and corrected clock bias and clock drift plots on (a), and estimated spoofer alterations on (b) for *third order attack* on Nexus 9.

### B. Real scenario results

This subsection presents results for real TEXBAT scenario 2 for both Nexus and SDR testbeds, and scenario 3 for the SDR.

#### 1) TEXBAT scenario 2 on Nexus 9

Fig. 7 shows the results for the real TEXBAT scenario 2 attack and mitigation with TSARM-S on the Nexus. An impressive 26.72 m RMSE is achieved even with OTA replay effects. The spikes depicted in the jerk plot in Fig. 7(b) distinctly correspond to the trapezoidal shape of the attacked clock drift. This is clearly a *third order attack*. The estimated clock drift is very similar to the one seen on the attacked clock drift plot in Fig. 7(a), and the 600 m clock bias attack reported in [28] is accurately estimated. The corrected clock bias and drift are maintained very closely to the clean version.

#### 2) TEXBAT scenario 2 on SDR

The SDR evaluation results for the real TEXBAT scenario 2 are seen in Fig. 8. The outputs are similar to the Nexus with a slightly higher RMSE of 31.83 m. The spikes are quite visible nonetheless, as seen in Fig. 8(b). A small deviation on the corrected clock bias of around 50 m is seen at the end of the plot. Further $\lambda$ tuning might improve such errors. The trapezoidal shape on the attacked clock drift is also seen in Fig. 8(a) as well as in the estimated spoofer clock drift in Fig 8(b).

#### 3) TEXBAT scenario 3 on SDR

Fig. 9 shows evaluation results for TEXBAT scenario 3 from the SDR. The attack on the clock bias of 600 m is clearly detected in the estimated spoofer alterations plot in Fig. 9(b). However, no significant spikes are seen in the clock jerk output. Nonetheless, The TSA was successfully detected and rejected based on a lack of measurement integrity for an all-channel simultaneous attack (it turns out that the scenario 3 attacks do not satisfy (6) or (7) for all channels). The proposed method achieved an RMSE of 15.92 m as a small ramp residual of 30 m on the corrected clock bias is seen in Fig. 9(a).

### C. Analysis and discussion

Table 5 shows the RMSE comparison summary between EKF and TSARM-S for all 9 scenarios. Overall, the SDR achieved an average 12.08 m clock bias RMSE, or roughly 40 ns time offset, outperforming the Nexus 9 with an average of 45.74 m RMSE, or 152 ns, for all tested scenarios. The SDR shows improved detection due to its stable clock bias and drift outputs and high configurability. The clock model on the SDR matches the Allan parameters more than the Nexus. Additionally, the Nexus OTA transmission adds further wireless channel effects. The proper clock modeling (Allan parameters) improves detection by handling the expected oscillator noise output, thus allowing more accurate behavior change detection in the clock jerk, as seen in the TSARM-S spoofer alteration outputs.

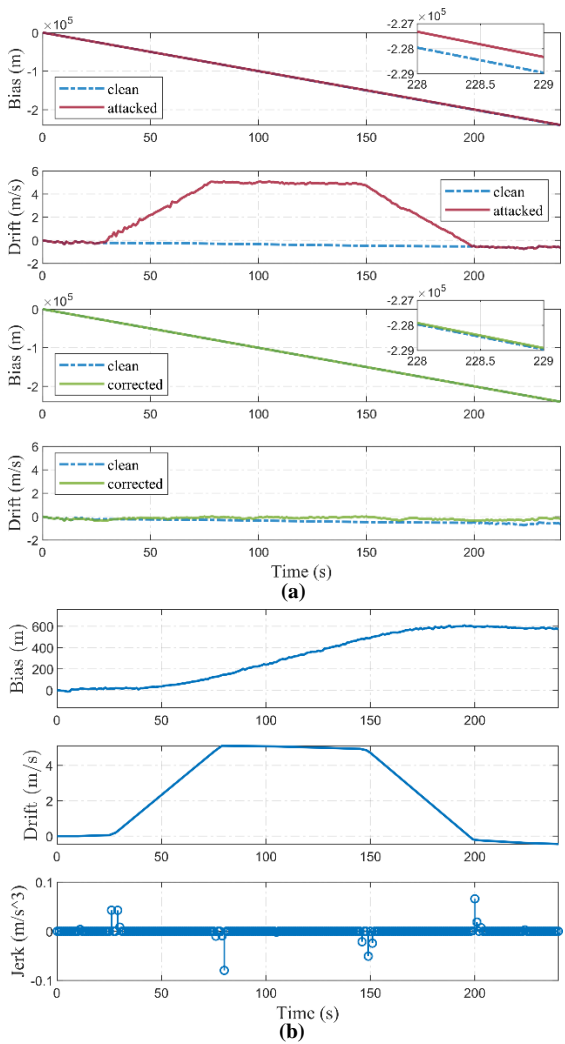

Fig. 7. Clean, attacked, and corrected clock bias and clock drift plots on (a), and estimated spoofer alterations on (b) for TEXBAT scenario 2 on Nexus 9.

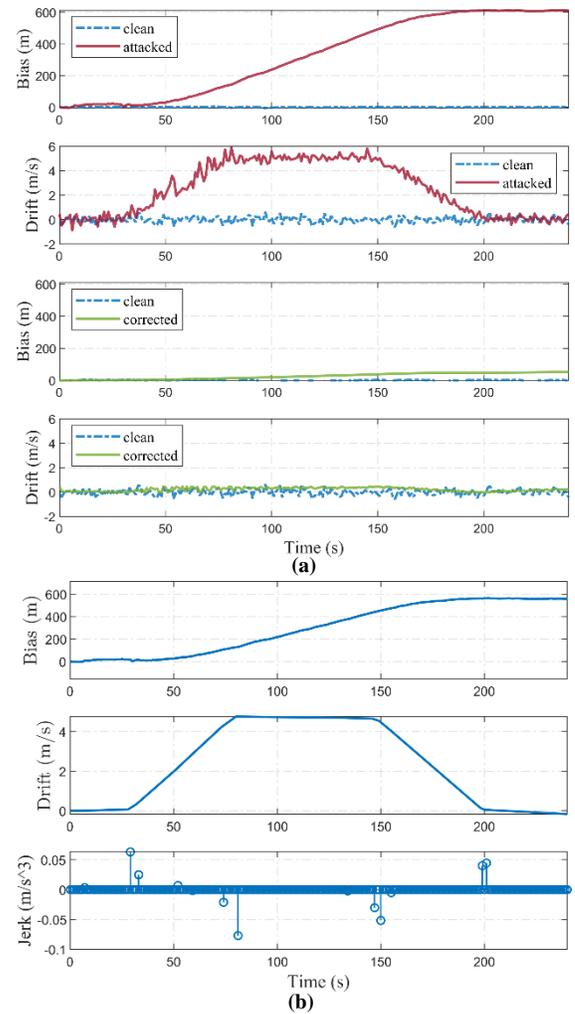

Fig. 8. Clean, attacked, and corrected clock bias and clock drift plots on (a), and estimated spoofer alterations on (b) for TEXBAT scenario 2 on SDR.

### 1) TEXBAT scenario 3 measurement integrity

As seen in Fig. 9, the TEXBAT scenario 3 lacks measurement integrity (see Section III-A). The clock drift does not follow the derivative of the clock bias. This can be seen as the attack on the clock bias reaches 600 m, while the drift remains unaltered in Fig. 9(a). Because the spoofer alterations are unconstrained in (12), the outlier variable captures these integrity discrepancies and successfully mitigates the attack on the clock bias. Thus, scenario 3 attack was successfully rejected due to a lack of measurement integrity and not because of a sparse event.

Also, the OTA experiments for this scenario with the Nexus testbed were unsuccessful: the device dropped channels during OTA replay and failed to attain a PVT solution after the attack had started around second 100. We hypothesize that the Nexus has simple self-integrity checks such as RAIM. In light of this evidence, the SDR was tuned to post-process scenario 3 in offline mode with deactivated features such as RAIM and other channel dropping mechanisms. Therefore, we conclude that a smart spoofer attack must maintain measurement integrity to delude rudimentary commercial device tests.

### D. Comparison with state-of-the-art techniques

In the following, TSARM-S is compared against a state-of-the-art TSA rejection technique, namely, Robust Estimator (RE) [16]. Similarly, a multipath sparse estimation (MPSE) technique is evaluated due to its similarity in terms of sparse detection [20]. The Nexus testbed along with synthetic *first order attack* is evaluated (see Table 3 and Fig. 1 for the attack). Fig. 10 shows the corrected clock bias (top) and drift (bottom) outputs from MPSE, RE, and TSARM-S against its clean version. The MPSE relies on an MP assumption that only a few pseudorange and pseudorange rate measurements will be affected. Since the TSA affects all measurements alike, the method is not effective in detecting the simulated *first order attack*. The RMSE for the MPSE is 1149.92 m. The RE achieves clock bias RMSE of 407.62 m. It is worth noting the RE's primary application is to reject TSAs that affect PMUs and therefore induce a clock bias error of 8000 m; thus a 1500 m attack is considered small.

### E. Application to a spoofing plus multipath scenario

To further validate that TSARM-S provides an accurate clock bias and drift under diverse settings, a scenario with both a TSA and MP is explored. Also, we compare with MPSE [20]. The TSA and MP scenario is evaluated on the Nexus testbed as a *first order attack* along with synthetic MP alterations as in [20]. Specifically, the MP is simulated similarly to a step attack with pseudorange alterations of 80 m and pseudorange rates of -24 m/s on two satellites. This MP is injected at similar times as the *first order attack*. Fig. 11 shows the clean and attacked clock bias and drift outputs of the EKF. The MP is barely noticeable in the bias since the TSA inflicts a 1500 m alteration, however the drift shows a step of around 9 m/s. Also, note that the MP model from [20] does not follow measurement integrity as seen in Section III-A. In the bottom two plots of Fig. 11, TSARM-S achieves an RMSE of 45.39 m, while MPSE achieves 1151.23 m RMSE. The bias seems unaffected by the MPSE as expected from TSA characteristics, however the drift seems corrected.

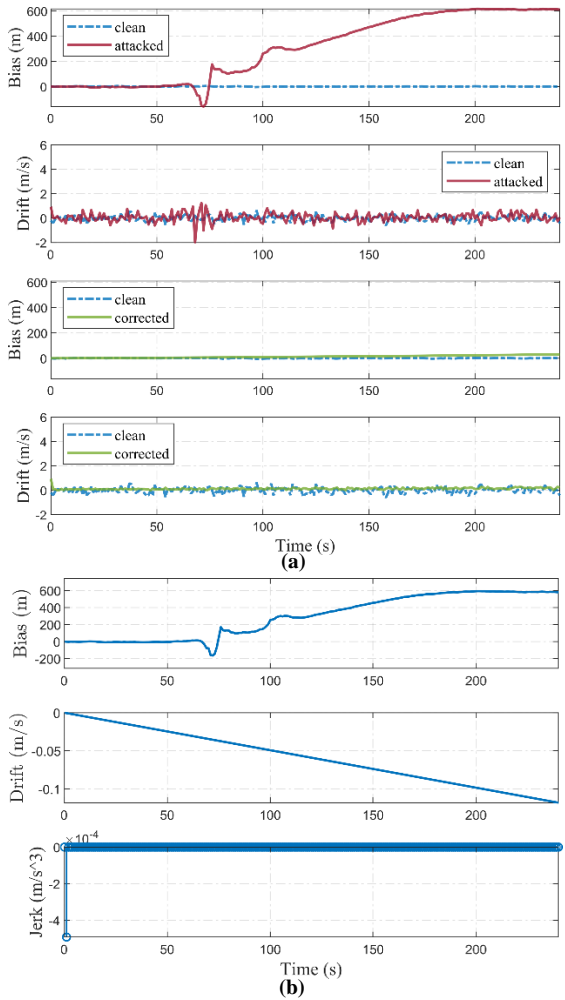

Fig. 9. Clean, attacked, and corrected clock bias and clock drift plots on (a), and estimated spoofer alterations on (b) for TEXBAT scenario 3 on SDR.

TABLE V
RMS ERROR RESULTS FOR SYNTHETIC AND REAL SCENARIOS (IN M)

|           | Nexus 9 | | SDR | |
|-----------|---------|---------|---------|---------|
| Scenario  | EKF     | TSARM-S | EKF     | TSARM-S |
| 1st order | 1151.40 | 20.17   | 1151.40 | 5.06    |
| 2nd order | 492.78  | 42.70   | 492.78  | 3.16    |
| 3rd order | 496.71  | 93.38   | 496.71  | 4.45    |
| Scenario 2| 425.39  | 26.72   | 403.91  | 31.83   |
| Scenario 3| N/A     | N/A     | 397.71  | 15.92   |

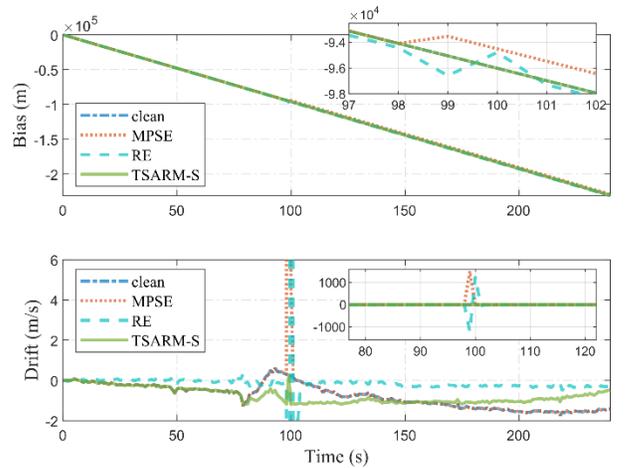

Fig. 10. Comparison output of clean and corrected clock bias and drift for state-of-the-art techniques with TSARM-S.

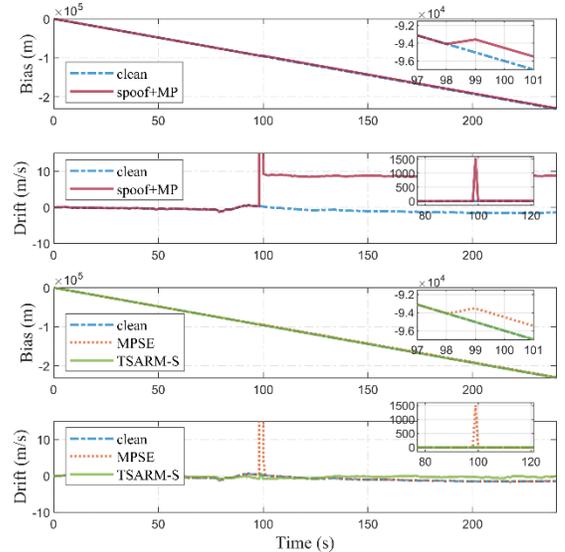

Fig. 11. Clean, attacked, and corrected clock bias and drift for a TSA and MP scenario.

Nonetheless, TSARM-S corrects the TSA within an acceptable error. It is worth emphasizing that TSARM-S is not intended for multipath mitigation, however this simulation numerically demonstrates that a reasonably accurate clock bias estimate can be produced, even if some multipath outliers are present.

## VII. CONCLUSION AND FUTURE WORK

This work presented a novel modeling of GPS TSA based on higher-order sparse-domains where the attack appears as a spike while a behavior change can be detected on the user clock. Further, it proposed a time synchronization attack rejection and mitigation technique based on a joint dynamic model and a higher order total variation operator. A test methodology was applied to first order, second order, and third order spoofing attacks as described in the TSA modeling. Also, real-data TEXBAT scenarios were evaluated. These attacks were successfully tested in a commercial receiver and a GPS SDR receiver [29]. In both testbeds, TSARM-S rejected smart spoofing attacks and achieved an average RMSE of 12.08 m for the SDR testbed, and 45.74 m for the Nexus. Both results translate to an RMSE of roughly 40 ns and 153 ns, respectively. TSARM-S was also evaluated against state-of-the-art spoofing mitigation techniques and MP techniques. Numerical simulations demonstrated that TSARM-S achieves reasonably accurate clock bias estimates under TSA and MP scenarios.

The proposed method is computationally feasible and can be implemented as an inexpensive firmware upgrade for commercial receivers. TSARM-S proved that a proper dynamic clock bias and drift model can achieve TSA rejection tasks by sensing behavior changes in higher-order sparse domains as well as measurement integrity gaps. Finally, due to the proposed systematic TSA modeling, simple adjustments to the dynamic model in (12) can potentially detect more complex (and smoother) attacks at further higher-order domains.

As future work, further improvement on the clock model (Allan coefficients) on the target device such as the Nexus 9 is planned to improve sensitivity in the sparse domain. Also, further tuning of the lambda parameter for different scenarios is expected. Finally, a real-time implementation of TSARM-S on the UTSA SDR is anticipated.